\documentclass[sigconf]{acmart}
\usepackage{enumitem}
\usepackage{hyperref}
\usepackage{wrapfig}
\AtBeginDocument{%
  \providecommand\BibTeX{{%
    \normalfont B\kern-0.5em{\scshape i\kern-0.25em b}\kern-0.8em\TeX}}}

\copyrightyear{2022} 
\acmYear{2022} 
\setcopyright{acmcopyright}\acmConference[Chinese CHI 2022]{The Tenth International Symposium of Chinese CHI}{October 22--23, 2022}{Guangzhou, China and Online, China}
\acmBooktitle{The Tenth International Symposium of Chinese CHI (Chinese CHI 2022), October 22--23, 2022, Guangzhou, China and Online, China}
\acmPrice{15.00}
\acmDOI{10.1145/3565698.3565765}
\acmISBN{978-1-4503-9869-5/22/10}

\begin{document}

\title[Co-design the Modeling Process for Efficient VFL via Visualization]{VFLens: Co-design the Modeling Process for Efficient Vertical Federated Learning via Visualization}


\author{Yun Tian}
\affiliation{%
  \institution{School of Information Science and Technology, ShanghaiTech University}
  \city{Shanghai}
  \country{China}}
\email{tianyun@shanghaitech.edu.cn}

\author{He Wang}
\affiliation{%
  \institution{School of Information Science and Technology, ShanghaiTech University}
  \city{Shanghai}
  \country{China}}
  \email{wanghe1@shanghaitech.edu.cn}

\author{Laixin Xie}
\affiliation{%
  \institution{School of Information Science and Technology, ShanghaiTech University}
  \city{Shanghai}
  \country{China}}
    \email{xielx@shanghaitech.edu.cn}

\author{Xiaojuan Ma}
\affiliation{%
  \institution{Department of Computer Science and Engineering, The Hong Kong University of Science and Technology}
  \city{Hong Kong}
  \country{China}}
  \email{mxj@cse.ust.hk}

\author{Quan Li}
\authornote{The corresponding author.}
\affiliation{%
  \institution{School of Information Science and Technology, ShanghaiTech University}
  \city{Shanghai}
  \country{China}}
\email{liquan@shanghaitech.edu.cn}


\begin{abstract}
As a decentralized training approach, federated learning enables multiple organizations to jointly train a model without exposing their private data. This work investigates vertical federated learning (VFL) to address scenarios where collaborating organizations have the same set of users but with different features, and only one party holds the labels. While VFL shows good performance, practitioners often face uncertainty when preparing non-transparent, internal/external features and samples for the VFL training phase. Moreover, to balance the prediction accuracy and the resource consumption of model inference, practitioners require to know which subset of prediction instances is genuinely needed to invoke the VFL model for inference. To this end, we co-design the VFL modeling process by proposing an interactive real-time visualization system, \textit{VFLens}, to help practitioners with feature engineering, sample selection, and inference. A usage scenario, a quantitative experiment, and expert feedback suggest that \textit{VFLens} helps practitioners boost VFL efficiency at a lower cost with sufficient confidence.
\end{abstract}


\maketitle

\section{Introduction}
\par
\begin{figure*}
  \centering
  \includegraphics[width=\linewidth]{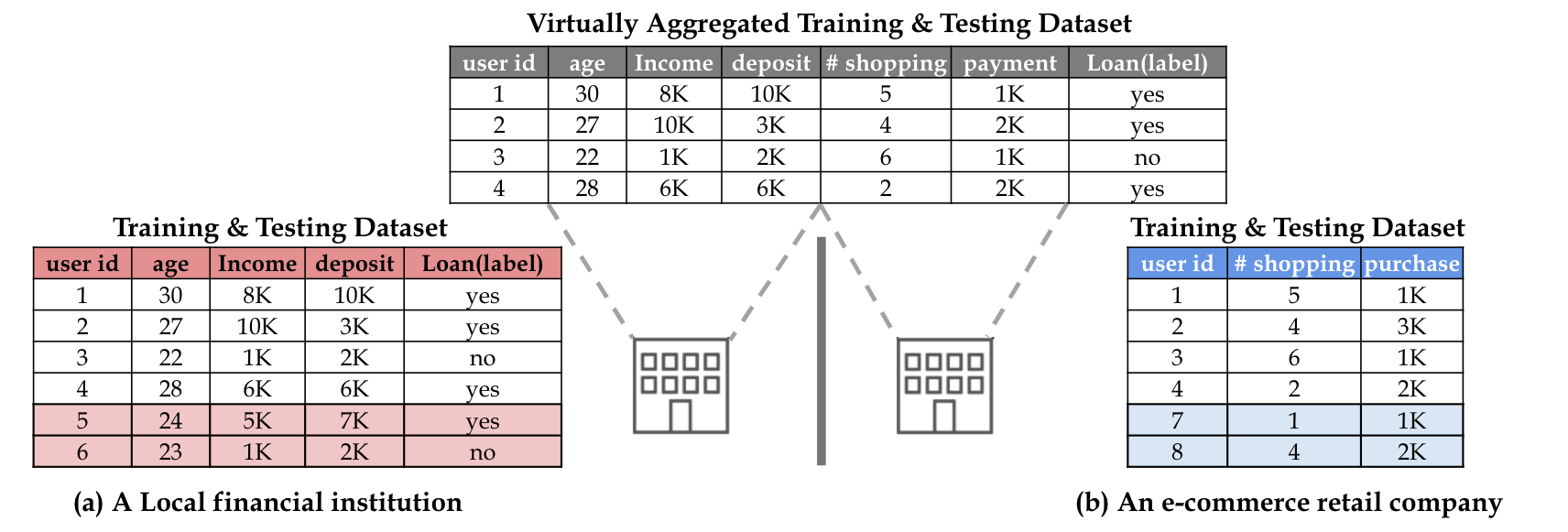}
  \vspace{-6mm}
  \caption{Suppose two parties, i.e., a local financial institution (a) and an e-commerce retail company (b) want to co-build a ML model for product purchase prediction. Only the financial institution has the label Y: loan or not and neither party wants to expose their features X. The two parties has an overlapping sample IDs, i.e., user id:$1$ -- $4$. The target is to establish a joint model under the condition of protecting privacy and the effect of the joint model is better than that of unilateral data modeling.}
  \label{fig:vfl}
\end{figure*}

 There is hope that all industries will benefit from big data-driven artificial intelligence (AI), especially after the huge success of \textit{AlphaGo}. However, with the exception of a few industries, most fields lack sufficient data or have data quality issues to support the construction of a reliable and robust machine learning (ML) model. At the same time, companies are reluctant to share or aggregate their valuable data in a centralized manner due to industry competition and privacy and security concerns, leaving the data often existing as a set of isolated data silos. As a viable decentralized solution that can potentially break down barriers between data sources while preserving privacy and security, federated learning (FL) enables users to collaboratively learn an ML model while keeping all data that may contain private information on their local device~\cite{BrendanMcMahan2017, yang2019federated}. Depending on how the data is partitioned between parties and application scenarios, FL can be divided into two main categories, namely horizontal FL (HFL) and vertical FL (VFL)~\cite{yang2019federated}. The focus of this study is VFL, also known as feature-based FL. VFL can be applied to situations where two datasets have considerable overlap in sample IDs but differ in feature space~\cite{cheng2021secureboost}. A typical example of VFL is a collaboration between an e-commerce retail company and a financial institution in the same city. Their customer set may contain the majority of residents in the area; therefore, the intersection of their customer spaces is huge. However, since the financial institution records its customers' income, spending behavior, and credit rating, while the e-commerce retailer retains its customers' browsing and purchasing history, their feature spaces are quite different. In this case, VFL allows both parties to train a joint ML model for product purchase prediction based on customer and product information under privacy-preserving security conditions (\autoref{fig:vfl}).

\par Although VFL has shown good performance in scenarios such as financial risk management~\cite{Fedai2019,Yang2018,zheng2020vertical}, healthcare~\cite{vepakomma2018split}, and e-commerce ad recommendation~\cite{yang2020federated}, real world practitioners have encountered the following challenges when trying to use VFL for their application domains~\cite{kairouz2019advances}: \textbf{1) Uncertainty in sample selection for training.} Traditional VFL practitioners mainly utilize two methods to prepare for the VFL training phase. First, when there is not much available data with labels, they may utilize all the overlapping samples with labels to train a joint VFL model for convenience. However, it is well known that the training speed of FL is much slower than that of the local model due to the design of data encryption and communication mechanisms. In some cases, utilizing all overlapping samples with labels for VFL model training can lead to much longer training time. Second, when the amount of data available for training is large, they may select some labeled data samples for training and evaluate the model based on general metrics such as \textit{accuracy}, \textit{loss}, and \textit{mAP}. As a slang expression in classical ML terminology, \textit``{garbage in, garbage out}''~\cite{hohman2020understanding} indicates the samples used for prediction should have a high-quality match with their specific jointly trained VFL models. Although the role of interactive data iteration in ML is emphasized and domain experts acknowledge that ``\textit{data samples need to have appropriate signals for the model to be useful}''~\cite{hohman2020understanding}, there is little support for their fine-tuning of training data samples in VFL scenarios. Both practices rely heavily on feedback from VFL model performance for further evaluation, which is sometimes too time-consuming and expensive. In particular, things get worse when the communication of the VFL training process is not so stable, as domain experts have to repeatedly re-upgrade the model training phase by trial and error. Therefore, an intuitive sample data evolution interaction mechanism that allows domain experts to compare the data characteristics and performance of different sample training datasets in a VFL scenario is necessary. \textbf{2) Non-transparent feature selection and assessment.} Successful ML applications require an iterative process to create models that provide the desired performance. One of the key processes involves feature engineering and in this study, we focus on feature selection and assessment. However, unlike traditional centralized ML modeling or HFL in which all data features are available and easy to assess, in VFL, participants update only their internal feature parameters during training, and external features from other parties are not visible to them due to the design of privacy-preserving mechanisms, which poses unique challenges for internal/external feature selection and assessment. That is, in addition to selecting the necessary internal features or transforming the original internal features into other powerful alternatives, practitioners are exploring how to assess external features from other parties~\cite{roschewitz2021ifedavg, wang2019interpret}. However, the lack of comprehensive consideration of the contribution of internal and external features while protecting privacy still undermines the use of VFL in production. \textbf{3) Costly and time-consuming inference.} The inference phase of VFL modeling requires online coordination between two (or more) parties to accomplish the inference task, which inevitably poses a challenge to computational resources and raises costs. According to our collaborating domain experts, the cost and deployment efficiency of federated modeling are issues that require rational planning for practical applications, and the use of homomorphic encryption in VFL can lead to a significant reduction in the computational speed and information transfer speed of federated modeling compared to centralized ML modeling~\cite{jing2019quantifying}. To solve this problem, in addition to optimizing the computational modeling process, another intuitive approach is to reduce the overall data volume. That is, not all samples to be predicted need to be truly predicted with the help of external features from other parties. For example, those samples in which practitioners have relatively high confidence in their labels do not need to be predicted by invoking an online trained VFL model because, e.g., the sample features are poor, and these samples can be safely ignored. Thus, how to visually help domain experts distinguish samples with different confidence in their labels is a desirable capability for real-world VFL deployments.

\par In this study, we co-design the modeling process to help VFL practitioners improve the efficiency of VFL modeling from the perspective of visualization. We first conduct an observational study of the current practices of collaborating domain experts to identify their main needs and concerns regarding VFL applications. Then, we streamline the analysis pipeline of feature and sample spaces and propose an interactive visualization system called \textit{VFLens}. \textit{VFLens} helps domain experts to interactively participate in feature selection, assessment, and sample data iteration processes before the VFL model training phase, in feature interpretation after the VFL model training phase, and in data sample selection during the VFL model inference phase. A case study and expert feedback confirm the efficacy of \textit{VFLens}. Our main contributions are summarized below.
\begin{itemize}
\item We describe the problem in the VFL context from the perspective of feature and sample space through an observational study and in-depth discussions of design requirements with VFL domain experts. 
\item We co-design the VFL modeling process to support domain experts to interactively participate in the data iteration, feature selection and assessment, and sample prediction processes. To the best of our knowledge, \textit{VFLens} is the first such effort in the VFL scenario. 
\item We evaluate \textit{VFLens} through a usage scenario, a quantitative experiment and expert interviews.
\end{itemize}

\section{Related Work}
\par The literature that overlaps this work can be categorized into four groups, namely, \textit{federated learning}, \textit{visualizations for federated learning}, \textit{feature selection and assessment}, and \textit{sample selection in machine learning}.

\subsection{Federated Learning}
\par Federated learning was first proposed by Google, which prevents data from being transmitted by distributing model training to each mobile terminal~\cite{BrendanMcMahan2017}. Later, they released the first commercial FL application, \textit{GBoard}~\cite{Hard2018}, which uses a recursive neural language model to predict the next word in a keyboard application. \textit{GBoard} allows each local mobile device to train the model using local data from the same distributed ML model. The global model can be updated by averaging the model parameters collected over all local models. Along the same lines, many studies have reshaped different ML models into a federated framework, including decision trees~\cite{li2020practical,zhao2018inprivate}, linear/logistic regression~\cite{li2018federated,mohri2019agnostic}, and neural networks~\cite{wang2020federated,yurochkin2019bayesian}. These works are categorized as HFL because the clients share the same feature space but differ in the sample space. Unlike HFL, VFL is applicable to scenarios where we have many overlapping instances but few overlapping features~\cite{Yang2018}. For example, an insurance company and an online retailer in a local city have many overlapping users, but each has its own feature space. VFL ``merges'' features and uses homomorphic encryption to protect the data privacy of the participating parties, and requires a more sophisticated mechanism to decompose the loss function of each party. This study focuses on VFL, ``virtually aggregation'' of different features to compute training losses and gradients in a privacy-preserving manner, and jointly build an ML model~\cite{cheng2021secureboost} with data from both parties.

\subsection{Visualizations for Federated Learning}
\par Researchers from academia and industry are using visualizations to demonstrate, explain, and monitor the process of federated learning. For example, in industry, Lenovo has simulated the industrial revolution in factories by demonstrating the process of horizontal federated learning to predict the internal pressure of hardware~\cite{Rojek2018}. Similarly, Cloudera Fast Forward Labs released an interactive simulation prototype, \textit{Turbofan Tycoon}, which takes advantage of visualization to examine the federated model and predict when a turbofan will fail~\cite{Mike2018}. \textit{FATEBoard}\footnote{https://fate.fedai.org/} utilizes dashboard visualizations to display modeling logs, metrics, and evaluation results, including information on data sets, job status, computational plots, and model output~\cite{Fan2018}. While \textit{FATEBoard} can help domain experts understand the ranking of features and the performance of models, it does not support detailed and interactive inspection of the sample and feature spaces. On the other hand, in academia, Wei et al.~\cite{Wei2019} developed a game to demonstrate the superiority of HFL and built a visualization prototype to help understand the operation of HFL. However, this work assumes that client-side data can be witnessed by the server-side. Li et al.~\cite{9408377} proposed \textit{HFLens}, which strictly follows a data privacy-preserving design and supports comparative visual interpretation at the overview, communication round, and client instance levels. \textit{HFLens} facilitates the investigation of the overall HFL process involving all clients, the correlation analysis of client information in one or different communication rounds, the identification of potential anomalies, and the evaluation of the contribution of each HFL client. However, the pain point for VFL is not the anomaly detection like \textit{HFLens}, because for VFL there are generally not as many data collaborators as for HFL, and the collaborators partnerships with common interests. In this work, we do not focus on the operational process of FL, but rather improve the efficiency of VFL modeling by involving domain experts in the sample and feature space.

\subsection{Feature Selection and Assessment}
\par There is a large amount of existing work related to feature selection~\cite{chandrashekar2014survey,blum1997selection}, which has two main difficulties. First, a large number of features are used in the process of building machine learning models; however, if several features are linearly correlated with each other, many of them will be redundant, which adds additional computational effort and leads to more complex parameters. Second, common feature analysis methods use feature correlation metrics, but correlation metrics cannot measure nonlinear relationships. Isabelle et al.~\cite{guyon2003introduction} performed a survey of automatic feature selection methods. The authors abstracted the core problem of feature selection, which is to find a minimal subset of features from a large number of features. The authors also argued that there are many options for feature selection and that there is no one universal and unique solution. There are other types of feature selection methods, such as wrappers~\cite{kohavi1997wrappers}, which iteratively eliminate features by regression or classification models to find the ideal subset of features. There are also metric-based methods~\cite{forman2003extensive,aphinyanaphongs2014comprehensive}, where users pick the top $k$ best features. However, they also suffer from the problems described earlier. As for feature assessment, it will be different in VFL modeling and traditional centralized machine learning modeling. In VFL, Host B does not have direct access to the features of Guest A, so in practice, the feature importance of Guest A is obtained by encrypting the IV values~\cite{chen2017fast,chen2018labeled,zhang2020batchcrypt} (Host B and Guest A will be introduced in \autoref{VFL Architecture}). In traditional centralized machine learning modeling, there are many methods to calculate feature importance, such as impurity-based feature importance~\cite{scornet2020trees} and permutation feature importance~\cite{altmann2010permutation}. In this study, we implement several different types of alternative feature selection techniques for choosing the internal features of Host B. We allow the user to decide whether the feature ranking is desirable or whether to focus on one of the features, while for the external features of Guest A, we use the encrypted IV values~\cite{chen2017fast,chen2018labeled,zhang2020batchcrypt} to obtain the feature importance metrics.

\subsection{Sample Selection in Machine Learning}
\par As one of the most critical infrastructures for building AI systems~\cite{halevy2009unreasonable}, data has a significant impact on the performance, fairness, robustness, and scalability of AI systems. However, data is often the ``\textit{the least motivated aspect, considered `operational'}'' relative to the lionized work in building new models and algorithms~\cite{halevy2009unreasonable,mehrabi2021survey,sambasivan2021everyone}. Akrong et al.~\cite{sambasivan2021everyone} reported on data practices of high-risk AI by interviewing AI practitioners around the world. They identified compound events of adverse and downstream effects caused by data problems named data cascades. Yee et al.~\cite{yee2003faceted} proposed Faceted browsing that allows users to use metadata to extract subsets of data that share desired attributes. The rank-by-feature framework allows users to examine low-dimensional projections of multidimensional data based on their statistics ~\cite{seo2005rank}. Hohman et al.~\cite{hohman2020understanding} proposed \textit{CHAMELEON}, which allows users to compare data features, training/test splits, and performance of multiple data versions. Facets\footnote{https://pair-code.github.io/facets/} helps developers examine ML datasets, including training/test segmentation, observe feature shapes, and explore individual observations. For data selection in FL, recent studies select relevant data distributively based on a benchmark model prior to training, regardless of other data quality factors or batch composition during training. Li et al.~\cite{li2021sample} provided a systematic analysis of the underlying data factors that affect FL model performance and propose an overall design to privately and efficiently select high quality data samples. However, the focus of these studies is on HFL. Inspired by their work, we emphasize the importance of training data in VFL and propose several interactive visualization schemes to facilitate sample selection prior to training of VFL models. In VFL inference, we separate the samples to be predicted based on the different confidence levels of their labels. To the best of our knowledge, \textit{VFLens} is the first attempt in this regard.

\section{Observational Study}
\subsection{Background}
\par
\begin{figure*}[h]
   \includegraphics[width=\linewidth]{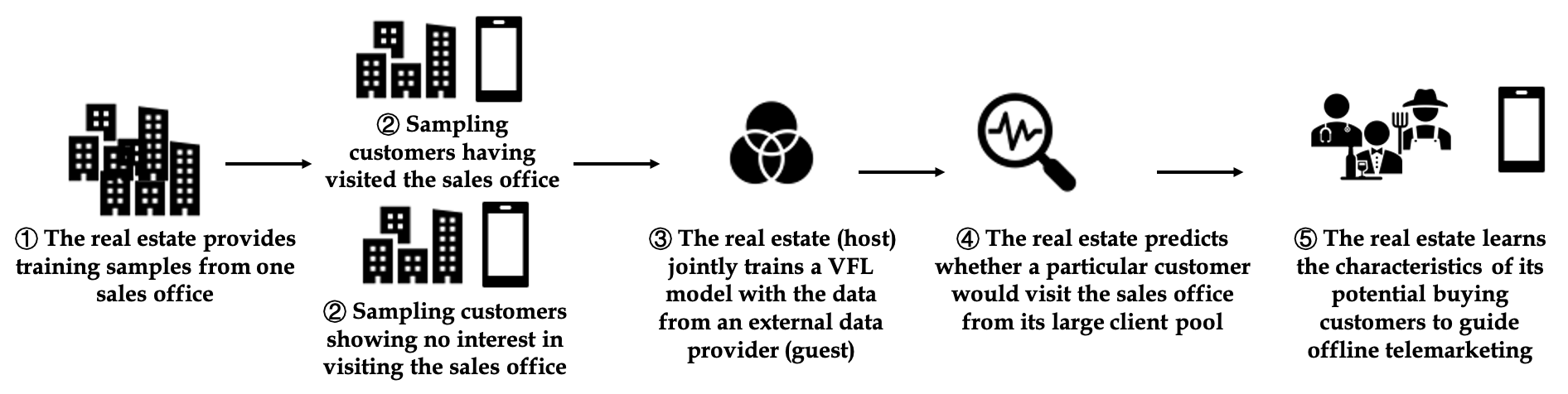}
   \vspace{-6mm}
  \caption{A case the experts encountered: designing and developing a precision marketing strategy in a real estate scenario. (1) The real estate company provides training samples from one sales office and (2) the samples are divided into two parts: those who have visited the sales office and those who have no interest in visiting the sales office. (3) The real estate company (host) jointly trains a VFL model with the data from an external data provider (guest). (4) The real estate company predicts whether a particular customer will visit the sales office from its large customer pool. (5) The real estate company learns the characteristics of its potential buying customers to guide offline telemarketing.}
  \label{fig:example}
\end{figure*}

 To understand the application of VFL in practice, we worked with a team of domain experts from a collaborating local financial and AI organization, including a FL project manager (E1, male, age: $31$), a VFL researcher (E2, male, age: $33$), two VFL engineers (E3, male, age: $27$, E4, male, age: $28$), and one business contact (E5, female, age: $29$). A large part of their work is to provide federated learning (both HFL and VFL) solutions to clients to meet their specific business needs. They shared with us a recent encounter in which they designed and developed a precision marketing strategy in a real estate scenario. Notably, the real estate company wanted to leverage the features of other parties through VFL to jointly train an ML model to predict whether a particular customer would come to visit the real estate sales office and understand the customer's characteristics. By taking this jointly trained VFL model, the real estate company can find more suitable customers from a large pool of other customers who may visit the sales office. An illustrative pipeline for this case is shown in \autoref{fig:example}.

\par In this case of the approval and start-up phases of VFL, the experts encountered several problems. First, the domain experts had to identify training samples from an overlapping set of users between the real estate (host) and the external data provider (guest, i.e., an online e-commercial and financial company). Although the total amount of available training samples has reached about $25,000$, not all of the samples are good enough. Most records were collected from field sales representatives and describe a rough profile of customer characteristics such as \textit``{gender}'', \textit``{age}'', \textit``{career}'', \textit``{income level}'', \textit``{marital status}'', and \textit``{family structure}''. E1 commented that ``\textit{there may be some outliers in the training samples because sometimes salepeople cannot guarantee the veracity of all characteristics.}'' E3 said that ``\textit{we are not experts in real estate,}'' so he did not have a clear idea about how to identify the right sample of customers for model training. Besides the local feature aspect, the limited computational resources and communication bandwidth between the host and the guest is another issue in the modeling process. In other words, dumping all overlapping training samples to train the VFL model is unrealistic because it may consume a lot of communication resources. Considering the characteristics of training samples and resources, domain experts would like to get some intuitive tips on how to select suitable and sufficient training samples. Second, when discussing how model knowledge can be used to guide their offline marketing strategies when contacting potential customers, E1 and E5 said, ``\textit{we should at least know what the model has learned so that we can understand the characteristics of customers who are likely to visit our real estate sales office.}'' Third, when it comes to inference and prediction using the trained VFL model, the initial expectation of the business requirement was that predictions needed to be made for all contacts in the real estate pool to obtain the likelihood that they would visit the sales office. However, the reality is that the pool is quite large, i.e., about $100,000$ and each prediction requires a paid call to the online VFL model. ``\textit{If the model predicts all the customers in the pool, the budget may not cover this cost.}'' said E4, ``\textit{can we just predict those customers who need the help of the VFL model without running all of them?}''

\subsection{Requirement Analysis}
\par To ensure that our approach was in line with the tasks and requirements, we interviewed all experts (E1 -- E5) to identify their main concerns about improving VFL modeling efficiency and have summarized their requirements below.

\par \textbf{R.1 Evaluate the quality of the training samples from the host.} The first pressing problem that experts encounter when building VFL models is to prepare sufficiently good training samples. While previous studies have proposed various methods to support data iteration for better model training~\cite{hohman2020understanding,sambasivan2021everyone,seo2005rank,yee2003faceted}, there is little support for this in VFL. E1 and E5 had little work experience in ML and, given the specific business requirements in the VFL scenario, they felt that their domain knowledge could be useful in selecting training samples. Therefore, both technologists (E3 and E4) and business personnel (E1 and E5) wanted to assess the quality of samples used by the host for model training in an intuitive and interactive way.

\par \textbf{R.2 Understand the internal/external features of hosts and guests.} Although the well-established automatic feature selection techniques allow analysts to confirm the contribution of each feature to the final prediction, especially in datasets with many features~\cite{chatzimparmpas2021featureenvi}, these techniques may produce significantly inconsistent results. According to E2, in a typical VFL scenario, the feature space is distributed in two (or multiple) parties. Thus, understanding internal/external features consists of two stages: 1) comparing alternative feature selection techniques based on their ranking of all internal features of the host, and 2) simulating external features of different batches of guests. Notably, the experts indicated that they would like answers to the following questions: ``\textit{which internal features are consistently ranked high?}''; ``\textit{How much does the technology vary in terms of feature ranking?}'' Our approach should allow experts to respond to such queries.

\par \textbf{R.3 Compare performance between models.} Inspired by recent studies that use similarities between model representations to correct for local training of the parties, such as conducting contrastive learning in model-level~\cite{li2021model} or comparing differences between the global HFL model and the local model~\cite{9408377}, E3 and E4 wished to understand the differences between each locally trained model and the global VFL model. For example, using standard validation metrics such as \textit{accuracy,} \textit{loss}, \textit{Kolmogorov-Smirnov (KS)}, \textit{Area under the curve (AUC)}, and \textit{mean Average Precision (mAP)} to understand performance fluctuations. In addition, experts wanted to have an overview of the history of the operations they performed so that they could identify ``\textit{critical points that might correspond to model performance improvements}''~\cite{li2021model}. Therefore, it is desirable to have an intuitive representation of the model performance per attempt and the performance differences between models.

\par \textbf{R.4 Obtain VFL forecasts at low cost.} The most critical business requirement raised by E1 and E5 is the conflict between the large number of forecasting requirements and the limited monetary budget and computational/communication resources. They commented, ``\textit{usually, the pool has a large number of samples to forecast, but blindly feeding all of them into the joint model will inevitably result in a waste of time and money.}'' Therefore, given the limited resources, experts need a strategy to balance the prediction sample size with a good enough prediction accuracy.

\par \textbf{R.5 Checking the prediction results.} As mentioned earlier, since our method makes a trade-off between the size of the VFL prediction sample and the prediction accuracy, experts are curious about the effectiveness of our strategy. Therefore, we should perform a comparative evaluation of the prediction results of our strategy. This evaluation will allow domain experts, especially E5, to understand the efficacy of our method and the reasons why those samples with relatively high confidence in the labels do not need to call the online trained VFL model for prediction.

\section{Co-design the Modeling Process of VFL}
\par
\begin{figure*}[h]
   \includegraphics[width=\linewidth]{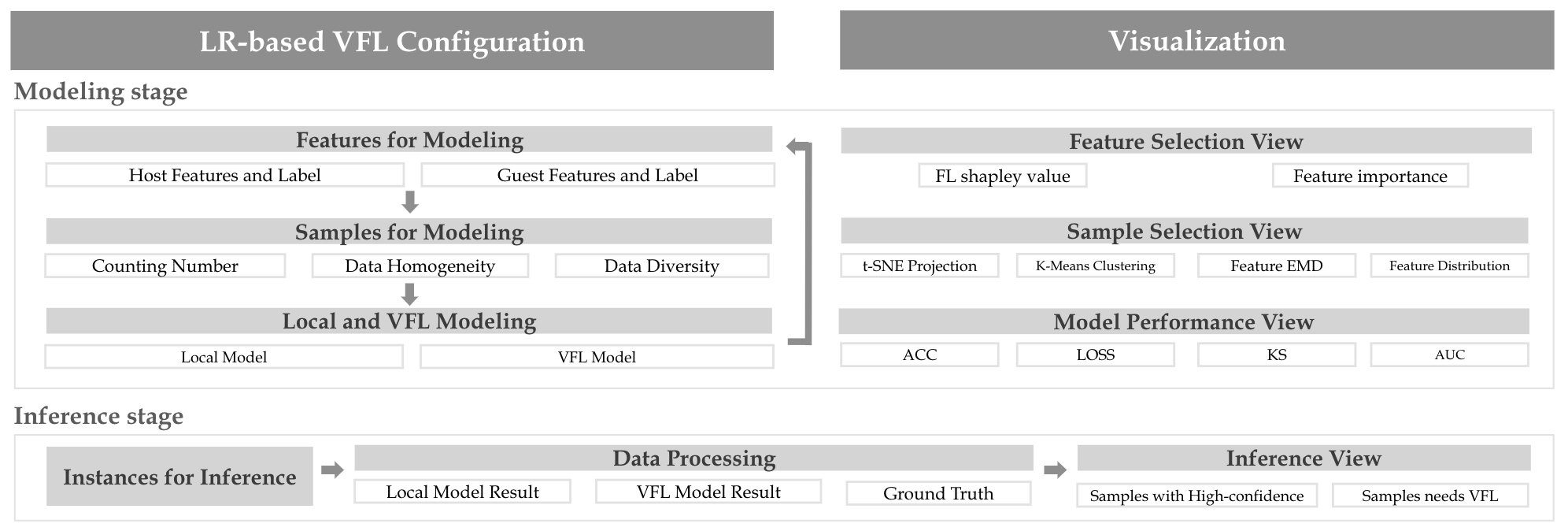}
      \vspace{-6mm}
  \caption{Overview of our approach. We divide the pipeline of our approach into two phases, namely \textit{modeling stage} and \textit{inference stage}. In the modeling stage, we first configure the backend using \textit{LR-based vertical federated learning}. Features distributed on two parties, i.e., host and guest, are selected and fine-tuned in \textit{Features for Modeling} and \textit{Feature Selection View}. After determining the features, we select the appropriate samples for modeling via \textit{Sample Selection View}. Then, we fine-tune the local model via \textit{Model Performance View} until we are satisfied and run the VFL model using the identified features and the samples initialized by the host side. Note that the local model can be fine-tuned several times. During the inference phase, all instances used for inference are compared and sampled via \textit{Inference View}. Finally, in \textit{Inference View}, samples with low confidence are fed into the VFL model for prediction; otherwise, samples with high confidence are provided to the local model for prediction.}
  \label{fig:pipeline}
\end{figure*}

 Inspired by the observational study, we propose a co-design process for VFL modeling, as shown in \autoref{fig:pipeline}. Our approach, named \textit{VFLens}, consists of an LR-based back-end VFL configuration module and a front-end visualization module. Specifically, the back-end module collects the necessary logs from the embedded VFL model and processes the information required for feature and data sample selection. The output of the back-end engine module is fed to the front-end visualization module for further analysis. The back-end engine module also receives interactive commands from the user through the front-end visualization module during the feature and sample selection phases of model training and prediction.

\subsection{VFL Architecture}
\label{VFL Architecture}
\par In this subsection, we illustrate the general architecture and basic background knowledge of a VFL system. According to the definition proposed by Yang et al.~\cite{yang2019federated}, VFL is suitable for scenarios with many overlapping instances but few overlapping features. For example, suppose two companies, $A$ and $B$, want to jointly train a machine learning model with their business data (i.e., they have different feature spaces). In addition, $B$ has the label data that the model needs to predict. In this case, $B$ is considered as the \textbf{Host} and $A$ as the \textbf{Guest}. Due to data privacy and security issues, the two companies cannot directly exchange their business data for training. To ensure data confidentiality during model training, an honest third party, usually played by an authority or a secure computing node, is introduced and participates without colluding with either party, i.e., \textbf{Collaborator} $C$. Both parties ($A$, $B$) are honest, but curious about each other's data. It is worth noting that a VFL training process usually consists of the two following phases as shown in \autoref{fig:architecture}(a), i.e., \textit{Encrypted Entity Alignment} and \textit{Encrypted Model Training}~\cite{yang2019federated}. In the \textit{Encrypted Entity Alignment} phase, VFL utilizes an encryption-based user ID alignment technique called Private Set Intersection (PSI)~\cite{huang2012private}, based on some encryption techniques such as \textit{MD5}~\cite{rivest1992md5} and \textit{Secure Hash Algorithm-1 (SHA-1)}~\cite{yuen2011chaos} to identify common users who overlap on both sides without exposing their respective data. Note that the VFL system does not tell non-overlapping users during the encrypted entity alignment process. Regarding the \textit{Encrypted Model Training} stage, we adopt logistic regression (LR)-based VFL~\cite{hardy2017private,yang2019quasi} to showcase our approach. Other federated privacy-preserving ML algorithms such as secure linear regression~\cite{yang2019federated} and SecureBoost~\cite{cheng2021secureboost} can also be quickly adopted or replaced with our back-end VFL solution. Since the initiator of VFL is the host, our proposed co-design process for VFL modeling (i.e., \textit{VFLens}) is oriented to the host side.

\subsection{LR-based VFL Configuration}
\par
\begin{figure*}[h]
    \includegraphics[width=\linewidth]{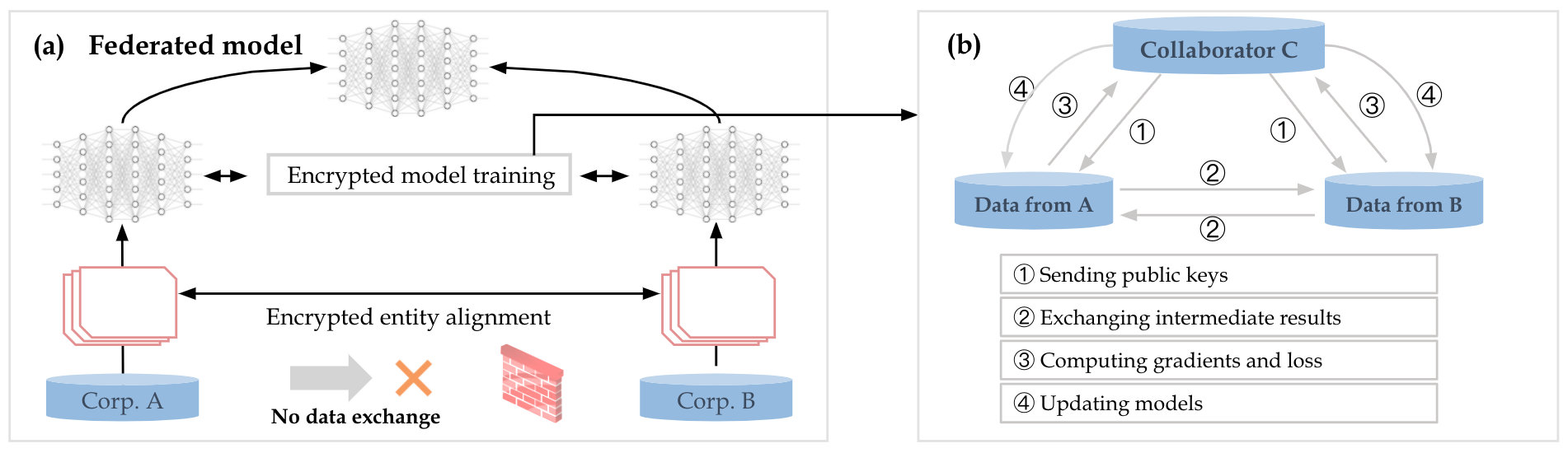}
         \vspace{-6mm}
  \caption{Architecture for a typical VFL system.}
  \label{fig:architecture}
\end{figure*}

 We utilize LR and stochastic gradient descent in cooperation with an additive homomorphic encryption scheme and mask~\cite{hardy2017private,yang2019quasi}. The phase of encryption model training starts after identifying overlapping entities common to both parties, which are used to train the ML model. Notably, our goal is to have both parties, i.e., the host and the guest, compute the intermediate results of the gradient separately as much as possible, and then get their gradient results through the interaction of encrypted information. As shown in \autoref{fig:architecture}(b) (1 -- 4), we divide the computational task as follows. That is, in each round of parameter update, each party needs to perform the following computations and interactions in turn and Step 1 -- 4 are repeated until the model converges.

\par \textbf{Step 1:} Guest A and Host B initialize their parameters, and Collaborator C generates a key pair and distributes the public key to A and B.
\par \textbf{Step 2:} Guest A computes its part of the gradient, encrypts with the public key, and sends it to Host B. Host B calculates its own amount of the gradient, encrypts it with the public key, and sends it to Guest A. All exchanged information is homomorphically encrypted, so parameters can be computed like in non-encrypted procedures, but are not visible. After receiving the corresponding parts, both parties compute their respective parts of the gradient separately.
\par \textbf{Step 3:} Both parties send the encrypted part of the gradient to Collaborator C for decryption, but to prevent Collaborator C from getting the gradient directly, Guest A and Host B add a random mask to the gradient part and send it to Collaborator C. Thus, the gradient obtained by Collaborator C cannot be used directly.
\par \textbf{Step 4:} Collaborator C gets the two parts of the encryption gradient, decrypts them, and returns them to Guest A and Host B, respectively. Then, Guest A and Host B subtract the previously added random mask to get the actual gradient and update their parameters.

\subsection{VFL Modeling Phase}
\par In this subsection, we first describe the general process of the VFL modeling phase. Then, we describe how \textit{VFLens} supports domain experts to co-design the feature and sample space for VFL modeling.

\subsubsection{General Process of VFL Modeling}
\par To train the VFL model, first, both host and guest collide de-identifiable\footnote{A process that removes personal identity and makes it impossible to identify or associate the subject of the personal information without additional information.} sample users id to determine the user intersection set. In the modeling phase, we decide which samples from the intersection will be used as training and testing samples (\textbf{\autoref{Samples for Modeling} Samples for Modeling}), where the training samples are used to train the VFL model and the testing samples are used to validate the model. Next, both parties train a ``semi-model'' using the features and labels of the samples. In addition to the samples, the model training also requires the determination of sample features (\textbf{\autoref{Features for Modeling} Features for Modeling}) and sample performance (sample labels). Specifically, the guest provides a certain amount of modeling sample features ($Feature_{guest}$) for VFL model training, and completes its ``semi-model'' training in its local environment. The host provides labels of the modeling samples for VFL model training, but it is not necessary to provide the features ($Feature_{host}$) of the modeling samples for VFL model training. Based on feedback from domain experts, ``\textit{host does not necessarily have the features of the modeling samples, but must have the labels of the modeling samples.}'' Without loss of generality, we assume that the host also has some features of the modeling samples, i.e., in our case, the host has both some but limited features and all the labels of the modeling samples. Thus, the host can also train a ``semi-model'' locally by using these features and labels.

\begin{figure*}[h]
   \includegraphics[width=\linewidth]{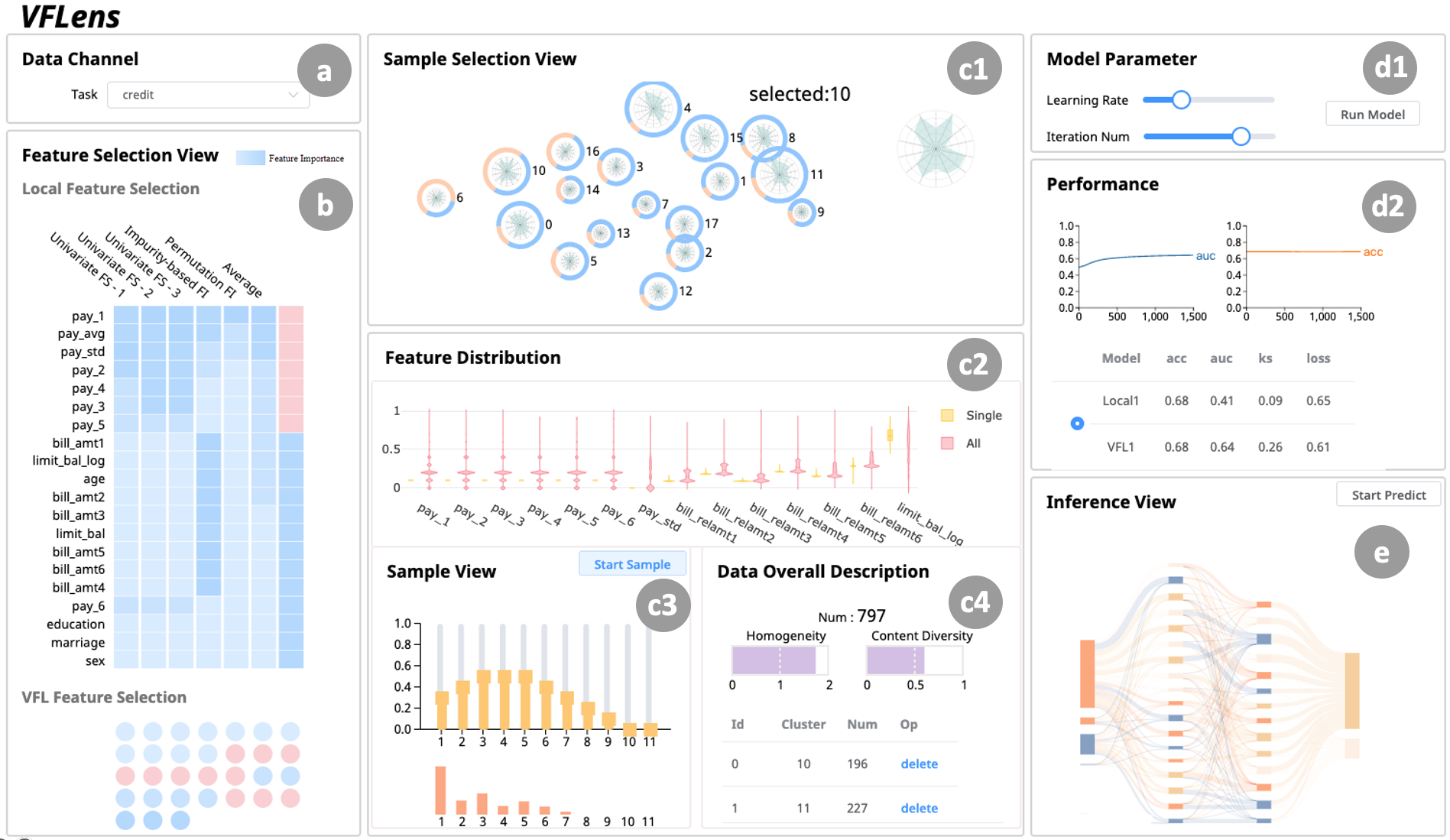}
   \vspace{-6mm}
   \caption{\textit{VFLens} interface: (a) A data loader for selecting cases of interest; (b) A feature selection view embeds five automatic feature selection methods. Users can select a certain number of local features and external ``invisible'' features by considering the internal feature importance and balancing the cost of VFL modeling. (c) The sample selection view shows the statistics information of the dataset and 2D embedding projection of all training samples. Users can select a particular data cluster in (c1) for selection and observe the detailed feature distribution of each cluster in (c2). (c3) An interactive scheme to select samples from the center to the periphery of a cluster. (c4) The quality of the training data set is evaluated by two metrics, \textit{homogeneity} and \textit{diversity}. (d1) Parameter selection for modeling. (d2) The model performance view for training the local and VFL model. (e) The summary view presents and visualizes the results of the classification results of all the samples to be predicted with our strategy.}
  \label{fig:system}
\end{figure*}

\subsubsection{Features for Modeling}
\label{Features for Modeling}
\par As mentioned earlier, we assume that the host holds user data that can be processed as features and labels. The host then needs to transform the user samples used for modeling into user features and user labels in order to complete the host's ``semi-model'' training locally. Therefore, how to properly co-design the host samples with features is of great interest to federated learning practitioners.

\par For internal features, we explore the feature importance-based host feature space~\cite{chatzimparmpas2021featureenvi} using the following five representative automatic feature selection techniques. There are three methods for \textit{Univariate Feature Selection}~\cite{jovic2015review}. 1) The first one uses the method of \textit{ANOVA} F-value test to select $k$ best features. To avoid automatic feature removal, we always set the value of $k$ to the maximum value of all features retained and let the domain experts decide which features to retain. 2) The $\mathcal{X}^2$-based method uses a chi-square test to select the $k$ best features. 3) The mutual information method estimates the mutual information of discrete target variables, with higher values implying stronger dependencies. 4) \textit{Impurity-based Feature Importance}~\cite{scornet2020trees}. This is related to the intrinsic nature of the ensemble algorithm, i.e., outputting feature importance after training. Therefore, we derive the feature importance from the best model found so far. 5) \textit{Permutation Feature Importance}~\cite{altmann2010permutation}. This is a technique for monitoring the reduction of model scores when individual feature values are randomly shuffled. To put them in the same context for comparison, we normalize the output from $0$ to $1$. We also calculate their average value to represent the feature performance on an average basis.

\par For external features, since they are invisible to the host, the data requester does not have access to the data information of the training participants in advance. The simplest way is that if the host wants to perform VFL jointly, they need to try all combinations of external features. This would waste a lot of training time. Existing work usually utilize feature bucketing~\cite{chen2017fast,chen2018labeled,zhang2020batchcrypt} for external feature engineering and secures the data to measure the importance of external features. In this work, we also utilize the cryptographic communication method of feature bucketing to compute the information values (iv) of external features.

\subsubsection{Feature Selection View}

\begin{figure}
 \centering
 \includegraphics[width=\columnwidth]{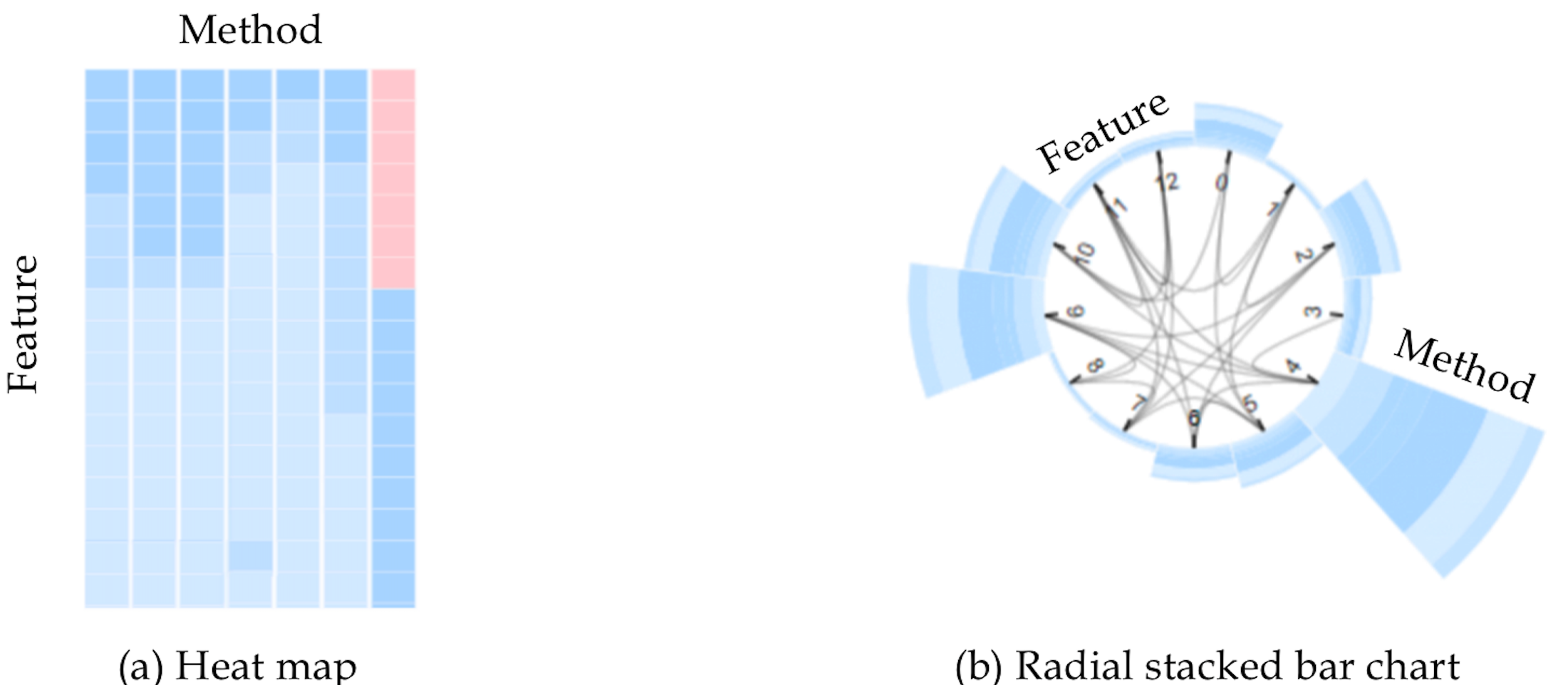}
 \vspace{-6mm}
 \caption{Design alternatives for feature selection.}
 \label{fig:feature_selection_view}
  \vspace{-3mm}
\end{figure}

\par Based on the above internal feature selection for modeling, we design a feature selection view \autoref{fig:feature_selection_view}(a) that supports domain experts to interactively select essential features from internal and external features based on feature importance (\textbf{R2}). The view is designed using a heatmap~\cite{chatzimparmpas2021featureenvi}, where the horizontal axis refers to different feature selection methods and the vertical axis represents different features. The color depth of the heatmap represents the importance of the features, and the penultimate column is the average of the first five methods, which serves as a reference for our selection. The last column can be interactively toggled to select and decide which local feature we finally choose for modeling. For external feature selection, due to the privacy mechanism of federated feature modeling, domain experts only have access to the importance of federated features and the anonymous ID of external feature. Therefore, \textit{VFLens} uses the depth of the color to encode the information values (iv) in the VFL feature selection view~\autoref{fig:system}(b).

\par We initially considered an alternative design solution displayed as a radial stacked bar chart~\autoref{fig:feature_selection_view}(b). We accumulate the scores of the different automatic feature selection methods in a stacked plot, which can directly display the most critical feature. In addition, thin lines connect the features, and their thickness represents the correlation between the two corresponding features. However, this initial design was not accepted by our collaboration experts when comparing the differences between the various feature selection methods. In addition, we need to consider the contribution of features from the external party. Therefore, we finally chose the first design.

\subsubsection{Samples for Modeling}
\label{Samples for Modeling}

\begin{figure}[h]
 \centering
 \includegraphics[scale = 0.2]{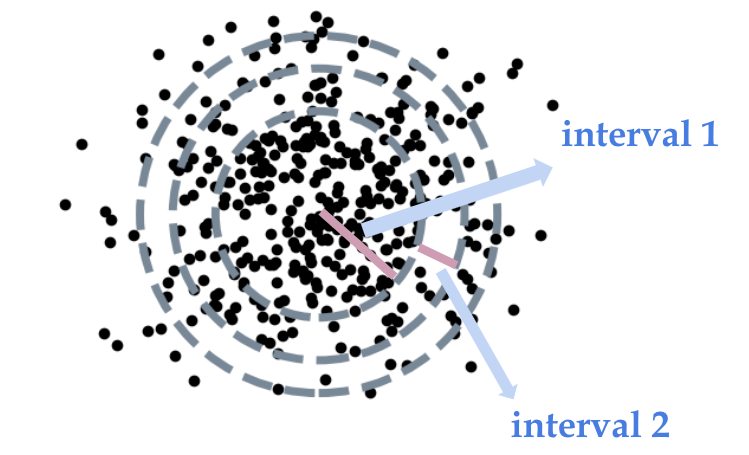}
 \vspace{-6mm}
 \caption{Illustration of interactive sampling stage.}
 \label{fig:sampling}
\end{figure}

\par In addition to assessing the contribution of features, another concern of VFL practitioners is to prepare suitable samples for the VFL modeling process. By selecting relevant samples, the labels of the samples can be potentially balanced (\textbf{R1}) and the diversity of sample features can be maintained, which can potentially improve model performance~\cite{li2006method}. To this end, we propose an interactive sampling and evaluation procedure, which consists of two phases, \textit{projection and clustering} and \textit{interactive sampling for each cluster}.

\par \textbf{Projection and Clustering.} We utilize \textit{t-SNE} as a dimensionality reduction technique because it is particularly good at conveying meaningful insights about the data, such as clusters and outliers~\cite{li2018embeddingvis}. Nevertheless, it may not be feasible to transform all avaiable samples in the intersection user set (i.e., the samples for training, validation, and prediction) into a two-dimensional representation based on the data features residing in the host and depict them on a two-dimensional space, as the data size may be huge. To deal with this potential problem, we abstract the raw training data into several clusters using \textit{KMeans++}~\cite{arthur2006k}, which augments \textit{KMeans} with a 
random and direct seeding technique that greatly improves its speed. We determine the number of clusters using the Elbow method~\cite{syakur2018integration}.

\par \textbf{Interactive Sampling.} Sampling is a simple but practically meaningful method to greatly reduce the data while maintaining certain data properties. We propose an interactive sampling scheme for each cluster. The basic principle is to drop most of the internal samples and keep the boundary samples to ensure strong learning and generalization~\cite{li2006method}. Users can interactively control the sampling ratio for a given data interval by adjusting the sampling slider in \autoref{fig:system}(c3). 
\autoref{fig:sampling} is a schematic of the sample data around the cluster, and the dashed circles are intervals divided from the origin outward, centered on the center of the cluster. The intuitive idea is that some nodes at the center of the cluster are more concentrated and representative, while sample data that are off-center tend to have some noise. We divide each cluster into intervals of $n$ from the center to the circle margin as $[0:\sqrt r]$,$[\sqrt r:\sqrt {2r}]$,$\dots $,$[\sqrt {(n-2)r}:\sqrt {(n-1)r}]$ and $[\sqrt {(n-1)r}:+\infty]$ with a constant $r$ just as \autoref{fig:sampling} (in this work, we define the value of $n$ as $11$, $r$ as $0.5$ after several experiments). This sampling scheme ensures that each individual region within the cluster occupies the same area. We sample the data according to a polar coordinates design because of the clustering mechanism of \textit{KMeans}, which clusters the samples according to the distance between the samples and the centroids. After completing the sampling of each cluster, we merge the sampled instances to the data pool.

\par \textbf{Data Sampling Evaluation.} We utilize \textit{target label balance} and \textit{feature diversity} as core metrics to help evaluate the combined data samples after interactive sampling. Regarding the target label of the training data, we find that the performance of the model deteriorates significantly as the imbalance increases~\cite{9488723}. When the classification distributions of the datasets are almost identical, the reduction in content diversity will lead to a considerable loss of accuracy. Thus, larger homogeneity and content diversity are likely to lead to better model performance. These two metrics are defined as follows. (1) \textit{Statistical Homogeneity.} Let $\mathcal{Y}$ is the set of target categories. Cluster $\mathcal{C}_{k}$ has a dataset $\mathcal{D}_{k}=\left\{\left(x_{k}, y_{k}\right)\right\}$, where each data $x_{k}$ has a label $y_{k}. \mathcal{D}_{k}$ follows a categorical distribution $q_{k}$. The uniform categorical distribution over $\mathcal{Y}$ is $q_{u}$. The statistical homogeneity of $\mathcal{D}_{k}$ is defined as ~\cite{zhao2018federated},
\begin{equation}
\mu_{k}=2-\sqrt{\sum_{y \in \mathcal{Y}}\left|q_{k}\left(y_{k}=y\right)-q_{u}\left(y_{u}=y\right)\right|^{2}},
\end{equation}
measuring the similarity between distributions $q_{k}$ and $q_{u}$ over $\mathcal{Y}$. (2) \textit{Content Diversity.} Given a dataset $\mathcal{D}$ having $M$ samples or $M$ sub-collections of samples and let $v_{i}$ be the flattened features vector of the $i^{th}$ sample or $i^{th}$ subcollection of samples, the similarity function of two vectors is $S\left(v_{i}, v_{j}\right)$. The content diversity of $\mathcal{D}$ is defined as ~\cite{wu2015hear},
\begin{equation}
\rho(\mathcal{D})=1-\frac{\sum_{i, j \in[M], i \neq j} 2 S\left(v_{i}, v_{j}\right)}{M(M-1)}
\end{equation}

\subsubsection{Sample Selection View}
\par We design a sample selection view to facilitate the mentioned interactive sampling for modeling. As shown in \autoref{fig:system}(c1), after projecting the clusters into a 2D space, a key issue is to convey the data characteristics of each cluster in terms of label distribution and salient features to facilitate the exploration of the data. Notably, we design a novel glyph to represent the data characteristic of each cluster. It consists of an inner radar chart and an outer ring. The data features are arranged in the clockwise direction of the glyph. The values on each axis of the radar chart represent the distance between the probability distribution of the data in this dimension of the cluster and the probability distribution of the entire training data, calculated using EMD distance (i.e., Wasserstein distance)~\cite{zhao2018federated}. The outer ring shows the label distribution, with yellow representing $1$ and the blue representing $0$, e.g., in the case of binary classification. When users click on a glyph in \autoref{fig:system}(c1), \textit{VFLens} displays its data features in \autoref{fig:system}(c2) as a reference for the subsequent sampling operation in \autoref{fig:system}(c3). It is worth noting that users can conduct an interactive sampling procedure for each cluster by filtering any number of samples from the center to the subcontinent (\textbf{R4}). In general, the principle of interactive sampling is to balance the data labels to diversify and enrich the characteristics of the samples. Therefore, we evaluate the overall data after sampling in \autoref{fig:system}(c4), involving \textit{sample count}, \textit{homogeneity}, \textit{diversity}.

\subsubsection{VFL/Local Modeling and Model Verification}
\par After identifying the features and samples to be modeled, we come to the VFL modeling stage. Multiple rounds of communication are required between host and guest to exchange encrypted gradients, and both parties transmit the following information to each other: 1) encrypted gradient information, i.e., the derivation of weights derived from sample features and labels; 2) de-identifiable sample ID, which are used to align the gradient information and transmit the convergence direction of the model. This process is a typical VFL, and the specific information interaction and communication process can be referred to the LR-based VFL configuration. Users can train the model by pressing the buttons in \autoref{fig:system}(d1). After VFL modeling, a VFL model is built at this point. In addition, since the host contains certain features and full sample labels, we can train a local model at the host side.

\par After building the initial VFL model, both parties verify the accuracy of the VFL model using de-identifiable user ID, features, and labels of the verification user set. For ease of  understanding, we describe the verification process as follows: 1) The host transmits the de-identified ID of the verification user set to the guest to verify whether the guest holds the data of the validation user set. If not, the verification process ends. 2) If the guest holds the verification user set, both parties start invoking their own ``semi-models'' to compute the corresponding prediction score. Specifically, the guest obtains the features of the verification user set from its local data and inputs them into the guest's semi-model to calculate the prediction score. Similarly, since the host has certain sample features, it can also calculate its prediction scores by feeding the sample features into the host's semi-model. Admittedly, in some cases, the host may have only sample labels without any feature dimensions, so there is no need to compute its prediction score. 3) After computing the prediction scores for both parties, the guest transmits the prediction score to the host, who aggregates the prediction score of both parties to obtain the final prediction result. 4) Finally, the host compares the prediction results with the ground truth labels and decides whether to fine-tune the model.

\subsubsection{Model Performance View}
\par In the model performance view, we display the results of the metrics used to evaluate the performance of the local and VFL models for each configuration (\textbf{R3}). We can iterate over the local model multiple times, since all communication and modeling processes occur locally. However, for the VFL model, we should iterate carefully if good enough performance is achieved.. Notably, as shown in \autoref{fig:system}(d2), this view shows the traditional metrics used to evaluate binary or multiclass classifications, such as \textit{accuracy}, \textit{loss}, \textit{KS}, and \textit{AUC}. Each iteration of the local and VFL models is recorded to facilitate the selection of the best model for the final model prediction.

\subsection{Inference Stage}
\par In the inference stage, the host can initiate a prediction based on its real-world requirement and invoke the ``semi-model'' of both parties for inference. Generally, the ``semi-model'' of both parties will give a prediction score, and the guest will send its prediction score to the host. The host will aggregate the prediction scores of both parties to get the final prediction result. Specifically, the vertical federated learning inference process is as follows: 1) the host de-identifies the target user ID to be predicted and transmits the de-identified IDs to the guest. After receiving the de-identified IDs, the guest queries whether it holds these corresponding features of these IDs. If the guest does not have these features, the prediction ends; 2) If the guest contains the features of these IDs, both parties invoke their respective ``semi-models'' to calculate the prediction scores. For example, the guest reaches the target user data locally, obtains the features of the target IDs, and then inputs the guest's ``semi-model'' to obtain the prediction scores. Similarly, if the host holds some user features, the host computes its prediction scores for the target user IDs; otherwise, if the host does not have user features, it does not need to compute the prediction scores; 3) The guest transmits the prediction scores of its ``semi-model'' to the host, and the host aggregates the two parts to obtain the final prediction result. At this point, the VFL inference stage is over and all samples that need to be predicted will get the prediction result by VFL inference.

\subsubsection{Instances for Inference and Sample Selection}
\par In reality, the number of samples to be predicted can be very large. In the inference phase of VFL, ``semi-models'' from both parties are required to cooperate online to complete the prediction, which poses a great challenge in terms of communication quality, modeling resources and economic costs. According to the expert feedback, VFL can improve the confidence of prediction results by expanding the feature space, but ``\textit{this is not suitable for all samples to be predicted.}'' It is worth noting that with limited budget cost, E1 and E5 want a less costly method to accomplish prediction for all samples to be predicted, while guaranteeing the quality of the prediction results to some extent. Therefore, we need to classify the samples to be predicted and send only those samples with high uncertainty to the VFL model for inference. And for the samples with high label confidence, we can use the local model of the host for prediction (\textbf{R4}).

\par For this purpose, we design the following sample selection schemes in the inference stage: 1) The sample ${ps}_i$ to be predicted is input into the local model of the host for inference, and the prediction result is obtained. Note that the local model here is different from the ``semi-model'' of the host. The local model only considers the data features residing on the host side and the data labels. 2) In the sample space, we compute the similarity between ${ps}_i$ and all training samples in terms of the host feature space, e.g., using the Euclidean distance to obtain the training sample ${ps}_i$ that is closest to ${ts}_i$. 3) For ${ts}_i$, determine its ground truth label, the local model's prediction result and the prediction results of the VFL model are consistent. We consider ${ps}_i$ a plausible sample if and only if the results of the three parts agree and are the same as the prediction results of the host local model for ${ps}_i$, where the prediction results of the host local model can be directly used as the final prediction results; otherwise, ${ps}_i$ is considered to have high uncertainty and needs to be predicted by the VFL model. At this time, all the samples to be predicted will be automatically divided into two categories. One category is samples with high credibility that do not require federated prediction, and the other category is samples that require further inference using the VFL model.

\subsubsection{Inference View}
\par As mentioned before, we obtained the two classes of samples to be predicted. To visualize the results and present the current classification of our strategy (\textbf{R4 -- R5}), we compute a many-to-many mapping based on the shortest distance from the predicted data to the training samples. In particular, as shown in \autoref{fig:system}(e), we use the \textit{t-SNE} and \textit{KMeans++} clustering methods to abstract the large number of samples to be predicted into the Sankey diagram. The flow from left to right in the Sankey diagram is the number of samples to be predicted. The leftmost column shows the eight label combinations of the training samples, corresponding to three cases (i.e., ground truth label, local model prediction result, and VFL model prediction results). The middle two columns correspond to the training data and the samples to be predicted, respectively. The last column lists the number of samples with high reliability and the number of samples that need to be predicted by the VFL model.

\subsection{Interaction Among the Views}
\par Rich interactions are integrated into \textit{VFLens} to build a low-cost VFL model. \textit{1) Sorting}. After the user clicks the average button in the feature selection view, the system will sort the average importance of all local features from highest to lowest. In the VFL feature selection view, we encode the importance of VFL feature as the color depth of the color block, and the color block color will change after the specified feature is selected. \textit{2) Hover on.} In the sample selection view, when the user selects a data cluster, a zoomed-in view of the cluster content is presented at the edge of the view, which helps the user to identify the specific feature patterns of different clusters. \textit{3) Parameter editing.} In the sample view, we adjust the scale of the samples in different regions by mediating the height of the bars with the mouse.

\section{Evaluation}
\subsection{Usage Scenario}
\begin{figure*}[h]
   \includegraphics[width=\linewidth]{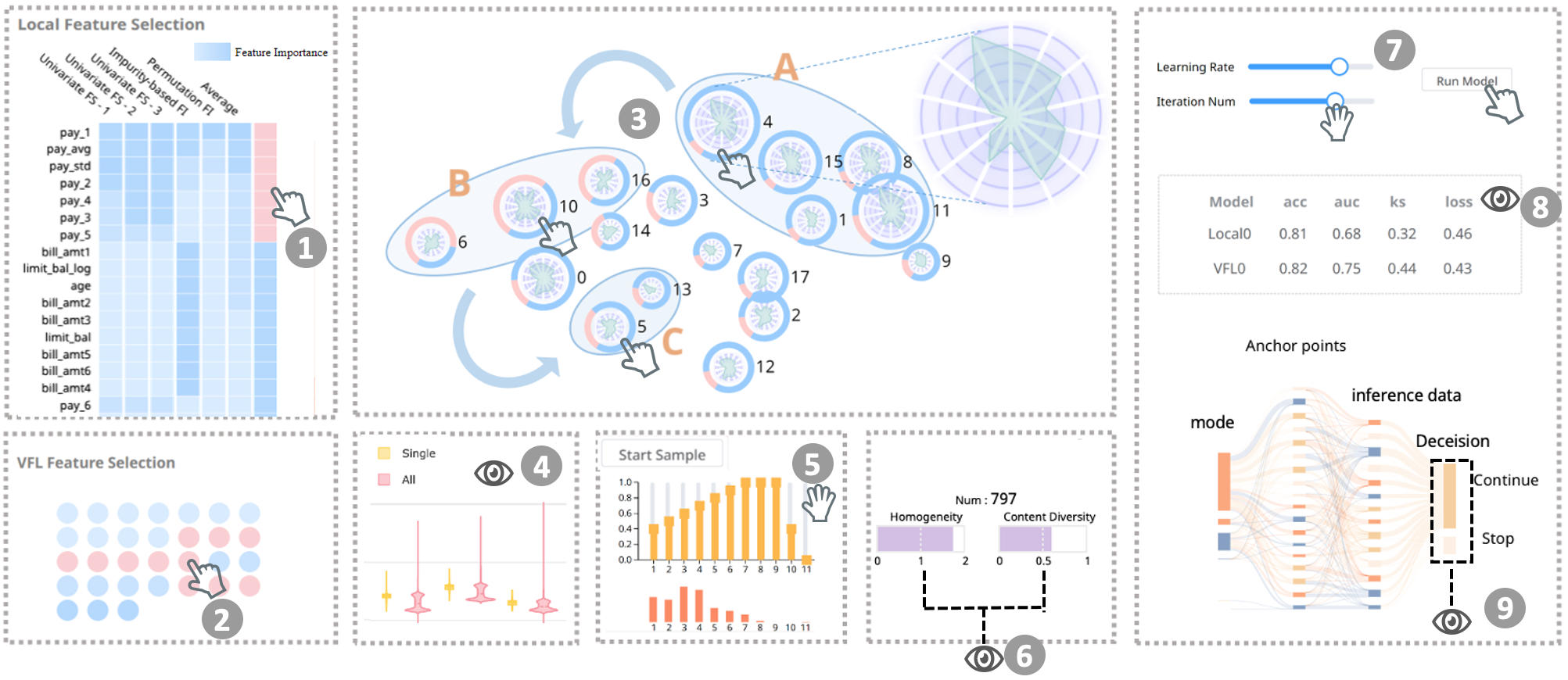}
   \vspace{-6mm}
  \caption{Usage Scenario: (1) Select the top-ranked local features in the local feature selection view. (2) External features with high iv values are selected in the VFL feature selection view. (3) In the sample selection view, we select different clusters for sampling based on the proportional of label in the cluster and the specific feature distribution information encoded in the clusters. (4) Observe the specific feature distribution information of the data in the cluster. (5) Set the sampling weight for different regions in each cluster. (6) Observe the two statistical indicators of the overall data that has been sampled and we  will stop sampling when the amount of data and the two statistical indicators reach the basic expectation value. (7) Adjust the learning parameters and start model training. (8) Run the local inference model, check the proportion of samples to be predicted, and verify the validity of the model.}
  \label{fig:case}
\end{figure*}

\par The modeling data from the real estate industry mentioned in the observational study has confidentially rules and cannot be used as experimental data. As an alternative, we take a publicly available credit card dataset~\cite{yeh2009comparisons}, where financial institutions can use their own customer data to predict various customer performance. Although most customer data are limited in the finance context, if other information can be used for joint modeling, financial institutions can obtain more accurate prediction models to further provide better services and reduce potential banking risks. To model this scenario, we use the dataset of customer defaults in Taiwan~\cite{yeh2009comparisons}. Notably, we use the first $14$ dimensions as internal features and the last $9$ dimensions as the federated external features. We also derive several alternative features. Finally, we obtain $20$ local features and $33$ external features for VFL modeling. In VFL, the host knows everything about the local features. Nevertheless, due to the privacy-preserving mechanism, we only use IDs to represent the federated external features. We use $15,000$ rows of total data in the VFL modeling stage and $15,000$ rows in the VFL inference stage.

\par \textbf{Feature Space Interaction.} After loading the local data into \textit{VFLens}, we select the top $13$ local features based on the recommended results of the standard automatic feature selection methods, as shown in \autoref{fig:case}(1). Regarding external features, to simulate reality, we randomly select $15$ remote external features for VFL modeling.

\par \textbf{Sample Space Interaction.} After the initial determination of the internal and external features, the host must select the appropriate number of samples for VFL model training. We turn to the sample selection view and first see a sharp feature in the upper left corner of the radar chart of cluster No.$4$. Based on the radar chart visualization, we find a similar distribution of features in clusters No.$1$, No.$8$, No. $15$, and No.$11$ next to cluster $4$. We then click on cluster No.$4$ and further observe the specific indicator of the radar chart enlarged on the right side of the view \autoref{fig:case}(3A) and its feature distribution in \autoref{fig:case}(4). We find that the feature distribution of this cluster has firm consistency. Therefore, we decide to reduce the overall sampling rate of these clusters to reduce the number of samples within these clusters for training, and to reduce the training cost while maintaining the loss of data features. More specifically, we first select cluster No. $4$, and in \autoref{fig:case}(4), we determine that the salient feature of this cluster is \textit{limit\_bal\_log}. The other features are similar to those of the overall samples. In \autoref{fig:case}(5), we reduce the number of duplicate training samples by reducing the sampling rate of the central region samples. Then we click the ``Start Sampling'' button of cluster No.$4$. The relevant information after sampling is displayed in \autoref{fig:case}(6). We can observe that \textit{Homogeneity} is $1.58$, which reflects that the balance of data labels is moderate. The metric of \textit{Content Diversity} is $0.10$, which indicates that the current sample features are relatively concentrated with a low degree of diversity. Following a similar procedure, we sample cluster No.$1$, cluster No.$8$, cluster No.$15$, and cluster No.$11$ at a lower sampling rate. The metric of \textit{Homogeneity} at this point becomes $1.58$, while \textit{Content Diversity} becomes $0.12$. We can see that the balance of the current samples is not significantly improved, since most of the target labels of these clusters are $0$. Inspired by this, we select clusters with more target labels of $1$, such as cluster No.$6$, cluster No.$10$, cluster No.$14$, and cluster No.$16$ in \autoref{fig:case}(3B). We first sample cluster No.$6$ and No.$10$ in appropriate proportions and observe that \textit{Homogeneity} changes to $1.77$ with significant improvement, while \textit{Content Diversity} (i.e., $0.21$), which also shows a huge improvement. However, we are still not satisfied with the value of \textit{Content Diversity}. We continue to operate on the cluster No.$13$ and No.$5$ in \autoref{fig:case}(3C) because they have more target labels. Finally, the \textit{Homogeneity} and \textit{Content Diversity} become $1.77$ and $0.64$, satisfying our modeling requirements.

\par \textbf{Modeling Training and Result Analysis.} Since the features and samples for training are ready, we come to the training phase, set the learning rate to $0.1$, and the number of iterations to $2,000$ in \autoref{fig:case}(7), and then perform VFL modeling. As shown in \autoref{fig:case}(8), we can see that the results of the training model are updated in real time for each iteration. Finally, the ACC of the local model is $0.68$, and the AUC is $0.60$. The vertical federated model has an ACC of $0.75$ and an AUC of $0.69$. The results show that our trained VFL model achieves better performance than the host's local model. However, the current AUC is still below $0.7$. We follow the above procedure to further improve the model performance, such as adding more external features required by the host and sampling other clusters for model training. Finally, the ACC of the local model reaches $0.78$ and its AUC reaches $0.64$, while the ACC of the VFL model is $0.79$ and its AUC is $0.75$.

\par \textbf{Model Inference.} After the model training, we then come to the model inference stage by clicking the ``start predict'' button. We observe that $6798$ records require further VFL inference, and the remaining $8202$ samples do not require calling the VFL model \autoref{fig:case}(9) (i.e., the rejection rate is $54.6$\%). To evaluate the efficacy of our interactive sampling strategy, we run the VFL model on the samples that do not need to invoke the VFL model to obtain their labels generated by the VFL model. We find that the results of federated inference for $7782$ samples are the same as the those of the local model (i.e., a hit rate of $94$\%), indicating that our interactive feature and sample selection strategy helps improve the efficacy of VFL modeling. As a comparison, we also use the VFL model to predict samples that require further VFL inference and find that the VFL model results for the $4707$ samples are the same as those of the local model. Compared to $94$\% hit rate for the unnecessary samples, $69.2$\% hit rate suggests that a larger proportion of samples is not confident in their labels. For these samples, we need to fine-tune more good features and train a better prediction model.

\subsection{Quantitative Experiment}
\par We also conduct quantitative experiments to compare the effect of \textit{VFLens} with other cases that do not use interactive feature selection and sample selection. As shown in \autoref{tab:experiments}, we can analyze that using only local data for modeling (Exp.1), the training will be very fast because no cryptographic operations are needed. However, the accuracy of the model will be very low due to the poor features. In Exp.2, all external features provided by the data provider are selected for federated modeling to obtain a better model. However, in real business scenarios, the cost spent on federated modeling is proportional to the sample size involved in the modeling process. So there is no need to select the entire aligned data for federated modeling. From the perspective of data reduction and reducing unnecessary costs, the corresponding (Exp.3 -- 4) is about \textit{VFLens}'s ability to use local feature selection and federated feature selection to eliminate some redundant features and also to reduce the tie of fine-tuning the model by domain experts. In Exp.5, we further use \textit{VFLens} for sample selection based on Exp.4, and we can see that the accuracy of the model is not significantly reduced, demonstrating the practical value of \textit{VFLens} in business scenarios.

\begin{table*}[h]
\centering
\caption{Training time(s) of local model and federated learning model on \textit{Default Credits Prediction} data in different situations.}
\label{tab:experiments}
\begin{tabular}{cccclll}
\hline 
Experiment id & Local feature  & FL feature   & Sample number      & AUC  & ACC   & \multicolumn{1}{c}{Time} \\ \hline 
1 & All(20/20)     & None(0/30)   & 10000         & 0.68 & 0.70   & \multicolumn{1}{c}{$<1$ min}             \\
2 & All(20/20)     & All(30/30)   & 10000         & 0.78 & 0.82    & \multicolumn{1}{c}{$>1$ day}           \\
3 & Part(13/20)    & All(30/30)   & 10000         & 0.78 & 0.80   & \multicolumn{1}{c}{$>12$ hours}          \\
4 & Part(13/20)    & Part(21/30)  & 10000          & 0.77 & 0.80   & \multicolumn{1}{c}{$>6$ hours}           \\
5 & Part(13/20)    & Part(21/30)  & 2100         & 0.75 & 0.79   & \multicolumn{1}{c}{2 hours}           \\
\hline  
\end{tabular}
\end{table*}

\section{Discussion and Limitation}
\par We conduct semi-structured interviews with all experts (E1 -- E5) to assess the efficacy of \textit{VFLens} and to determine whether our approach could help them improve the efficiency of VFL modeling.

\par \textbf{System Performance.} All experts agreed that \textit{VFLens} demonstrates a straightforward data science process with an intuitive visual design and practical implications. ``\textit{The system is very useful for hosts to assess the quality of their data and to further evaluate how much resources are needed for VFL modeling,}'' said E1. Our system has the potential to help a variety of commercial enterprises and sectors build economically usable VFL models. Considering the balance between resource input and model accuracy, the results seem to be quite good. \textit{VFLens} is the first system that attempts to address realistic pain points across business sectors, which will go a long way to ``\textit{enrich and complement existing federated learning frameworks}''. In discussing with our collaborating experts what inspired them most about our system, E2 said that they were most impressed with the fine-tuning and interactive sampling of local data features based on dimensionality reduction results, which ``\textit{fits our sense and it was very valuable.}''

\par \textbf{Visual Design and Learnability.} We draw from observational studies of experts working routinely in real-world scenario to inform the system design and classicial interfaces used in commercial federated learning products (e.g., \textit{FateBoard} in \textit{FATE}~\cite{Fan2018}). For example, the matrix design visually presents the feature importance provided by different automatic feature selection algorithms. The model performance represented by various metrics and the sample distribution represented by the histogram design are familiar visual metaphors in traditional data science pipelines and analysis. Experts commented, ``\textit{we can quickly get used to visual encoding because the interactions are the same as in traditional modeling.}'' Specifically, after a briefly introduction to each view and its capabilities, the experts developed customized exploration paths for interactive VFL modeling and inference.

\par \textbf{Generalizability and Scalability.} In this work, we only show our pipeline using LR-based VFL. Tree-based models such as \textit{XGBoost}~\cite{chen2016xgboost} and \textit{Random Forest}~\cite{ho1995random} are more commonly used in various commercial domains. Incorporating such models into \textit{VFLens} is not a difficult task. Nevertheless, the visualization design of the front-end may require unavoidable alternations, as the underlying models may introduce the necessary specific information to be checked. Notably, if other types of models or even deep learning models are used, the deployment and implementation of VFL versions of the models and a detailed discussion of model performance comparisons are major concerns for future work.

\par \textbf{Contributions Over Previous Work.} Compared to previous works, our system has inspired experts in many ways. They commented that when they utilize VFL in real-world scenarios, they often experience poor data quality or lack of sufficient data attributes when a large enterprise group consists of different affiliated business units that are unable to share data. Some business units even lack data labels. All of these real-world problems pose a huge challenge for VFL modeling. Experts said that if they want to leverage external data sources from the Guest using VFL, \textit{VFLens} can greatly help them sort out the current data quality before modeling. With an understanding of overall data and feature quality, experts can better understand whether they need to call on external data sources for VFL modeling. ``\textit{VFLens could help business units in companies assess data quality, such as data fill rate, label fill rate,}'' said E1, adding that this needs to be done before FL modeling. Also, \textit{VFLens} can help experts understand data samples and features before modeling, which has high application value in real-world scenarios.

\par \textbf{Limitation.} Our work has several limitations. First, interactive samplings of predicted clusters relies heavily on user's experience and domain knowledge, introducing uncertainty in sample selection. We should provide more intelligent guideline to help users perform interactive sampling more efficiently and confidently. Second, we only utilize representative metrics such as \textit{accuracy} and \textit{AUC} to compare different versions of the model. We did not consider using the internal information of the models themselves, such as gradient distribution and weight information, which may be more critical when employing tree-based or deep learning models.

\section{Conclusion and Future work}
\par In this study, we present \textit{VFLens}, a visual analytics system for interactive VFL modeling and inference, which improves VFL modeling efficiency by supporting domain experts to co-design internal features and interactively manipulate sample spaces. A usage scenario, a quantitative experiment, and expert feedback confirm the efficacy of \textit{VFLens}. In the future, we will combine more models with the FL version to enable experts to achieve better prediction results. Also, we will introduce more sampling methods to reduce the impact of anomalous samples on training and further discuss the design space when involving other types of models, such as tree-based or deep learning-based models.

\begin{acks}
We are grateful for the valuable feedback and comments provided by the anonymous reviewers. This work is partially supported by the Research Start-up Fund of ShanghaiTech University and HKUST-WeBank Joint Laboratory Project Grant No.: WEB19EG01-d.
\end{acks}

\bibliographystyle{ACM-Reference-Format}
\bibliography{sample-base}










\end{document}